\documentclass[apjl]{emulateapj}
\usepackage{graphicx}
\usepackage{amssymb}
\usepackage{array}

\shortauthors{Pu et al.}

\begin{document}

\title{TWO TYPES OF ERGOSPHERIC JETS FROM ACCRETING BLACK HOLES: THE DICHOTOMY OF FANAROFF-RILEY GALAXIES}

\author{Hung-Yi Pu\altaffilmark{1}, Kouichi Hirotani\altaffilmark{2}, Yosuke Mizuno\altaffilmark{3},  and Hsiang-Kuang Chang\altaffilmark{1,3}}

\altaffiltext{1}{Department of Physics, National Tsing Hua University, Hsinchu
30013, Taiwan}
\altaffiltext{2}{Theoretical Institute for Advanced Research in Astrophysics,
Academia Sinica, Institute of Astronomy and Astrophysics, P.O. Box
23-141, Taipei, Taiwan}
\altaffiltext{3}{Institute of Astronomy, National Tsing Hua University, Hsinchu
30013, Taiwan}

\begin{abstract}
We investigate  the extraction of the rotational energy of a black hole under different accreting environment.
When the accretion rate is moderate, the accretion disk consists of an outer thin disk and an inner advection-dominated
accretion flow. In such a combined disk, the outer thin disk can sustain a magnetic field with moderate strength at
the event horizon, leading to the formation of relativistic jets with moderate luminosity and speed via the 
magnetohrodynamic Penrose process. When the accretion rate increases enough, on the other hand, the disk becomes geometrically thin near the horizon. In this slim disk, the denser plasmas can sustain a stronger magnetic field than that
in a 
combined disk, leading to the formation of jets with greater luminosity and speed via the Blandford-Znajek processs. It 
is discussed that the former jets are associated with the Fanaroff-Riley (FR) I galaxies and the latter with FR II galaxies.

\end{abstract}


%\clearpage
\section{Introduction}
Remarkable advances have been made
in explaining the observed spectrum of black hole  X-ray binaries (BHXBs)
and active galactic nuclei (AGNs) with the help of theoretical studies on black hole (BH) accretion disks \citep{abr99,kat08}. 
Current stable BH accretion disk solutions include the advection-dominated accretion flow, or ADAF for short \citep{nar94,nar95,nar97}, the thin disk \citep{sha73,nov73}, and the slim disk \citep{abr88,abr10,sad09,sad11}.
The accretion disk type is a function of the the dimensionless accretion rate, $\dot{m}\equiv\dot{M}/\dot{M_{\rm Edd}}=L/L_{\rm Edd}$,
where $\dot{M}$ is the accretion rate, $\dot{M}_{\mathrm{Edd}}=L_{\mathrm{Edd}}/\xi c^{2}$ is
the Eddingtion accretion rate, $L_{\mathrm{Edd}}$
is the Eddingtion luminosity, $\xi$ is the efficiency ($\xi=1$ is adopted for later computation) and $c$ is the speed of light. 
In Table 1 we summarize the properties of the above disk solutions.

The accretion disk solution is also a function of radius \citep{abr95,che95}. A combined disk which consists an outer thin disk and an inner ADAF has been also proposed to explain observations of BHXBs and AGNs \citep{nar96,esi97,fen04,tru11}. Such disk is expected to be form before the entire disk transits from an ADAF to a thin disk, because the hot ADAF would cools down from outside. In other words, the transition radius from outer thin disk to inner ADAF solution decreases as the accretion rate increases until the entire disk become thin disk \citep{hon96,kat98,man00}. 
For  a comprehensive consideration of how accretion disk type vary with $\dot{m}$, we also include the the combined disk in Table 1.

\begin{table*}[ht]
\begin{center}
\label{table1}
\caption{Summary of Black Hole Disk Properties}
\begin{tabular}{>{\centering}b{2cm}>{\centering}m{4.2cm}>{\centering}m{2.cm}>{\centering}m{4.2cm}>{\centering}m{4.2cm}}
\hline
\hline  
Disk Type & ADAF& Combined Disk & Thin Disk & Slim Disk\tabularnewline
% & \multicolumn{2}{c}{\hfill (outer thin disk + inner ADAF)}    & 
%\tabularnewline
\hline 
 &  &  & \tabularnewline
Accretion Rate & $\dot{m}\ll0.01$ & $\dot{m}\lesssim0.01$ & $0.01\lesssim\dot{m}\lesssim0.3$ & $0.3\lesssim\dot{m}$\tabularnewline
 &  &  & \tabularnewline
Cooling Process & advection  & & radiation & advection\tabularnewline
 &  &  & \tabularnewline
Optical Depth & thin  & & thick & thick\tabularnewline
 &  &  & \tabularnewline
Disk Geometry & thick $(H\sim R)^{a}$ & outer thin disk + inner ADAF & thin $(H\ll R)$ & thick $(H\lesssim R)$\tabularnewline
 &  &  & \tabularnewline
%Global Geometry & thick & outer thin disk + inner ADAF & thin & slim\tabularnewline
%&  &  & \tabularnewline
Viscosity$^{b}$ & $\alpha\sim0.1$& & $\alpha\sim0.01$ & $\alpha\ll1$\tabularnewline
&  &  & \tabularnewline
Infalling Process & viscous-driven ($\ell_{\rm{in}}<\ell_{\mathrm{ms}}$)$^{c}$& &  viscous-driven ($\ell_{\rm{in}}\sim \ell{}_{\mathrm{ms}}$)  &  pressure-driven ($\ell_{\rm{in}}> \ell{}_{\mathrm{ms}}$)\tabularnewline
&  &  & \tabularnewline
Flow Geometry near the Horizon$^{d}$ & quasi-spherical& &quasi-spherical & disk-like\tabularnewline
&  &  & \tabularnewline
%Illustration Plot$^{e}$ & \plotone{ADAF.ps}& &\plotone{thin.ps} & \plotone{slim.ps}\tabularnewline
%&  &  & \tabularnewline
%Representative Angular Momentum Profile$^{f}$ & \plotone{ADAF_m.ps}& &\plotone{thin_m.ps} & \plotone{slim_m.ps}\tabularnewline
&  &  & \tabularnewline
%Relativistic Jet? &  & EJC & & EJS\tabularnewline
%&  &  & \tabularnewline

\hline
\end{tabular}

\end{center}

{$^a$} {$H$ is the disk height and $R$ is the radial distance to the central BH.}

{$^b$} {$\alpha$ is the parameter in $\alpha$-prescription \citep{sha73}.}
  
{$^c$} {$\ell_{\rm{in}}$ is the angular momentum of the flow when it falls onto the BH. See also Figure 1.} 

{$^d$} {Note that the geometry of the accretion flow {\em near the horizon} 
is determined by the `infalling process' (see text), therefore it
can be different from the `disk geometry', which describes the geometry of an accretion disk {\em far from the horizon}. See also Figure 1.}

%{$^e$} {Dominating force are shown: $F_{\rm g}$, force due to the gravity of the BH;  $F_{\rm c}$, force due to the rotation of the disk; $F_{\rm p}$, force due to the pressure gradient inside the disk. The dashed line represents the location of $R_{\rm ISCO}$.}

%{$^f$} {The Keplerian value is shown in blue dashed curve and the disk value is shown in red solid curve.}

\end{table*}

%Viscosity$^{b}$ & \multicolumn{2}{c}{$\alpha\sim0.2$} & $\alpha\sim0.01$ & $\alpha\ll1$\tabularnewline

As a pure relativistic effect, the specific angular momentum of a Keplerian rotating object near the BH, $\ell_{\rm k}$, has a local minimum at the radius of the innermost stable circular orbit, $R_{\rm ISCO}$.
The flow geometry {\em near the BH horizon} is determined by how
the angular momentum distribution of the accretion disk, $\ell(R)$, deviates from the Keplerian values \citep{abr98}. 
If  $\ell(R)>\ell_{\rm ms}$, where $\ell_{\rm ms}$ is the Keplerian augular momentum at $R_{\rm ISCO}$,
$\ell(R)$ \textit{must} equal the Keplerian value twice (assuming $\ell(R)=\ell_{\rm k}$ at $R_{1}$ and $R_{2}$, and $R_{2}>R_{\rm ISCO}>R_{1}$).  
In this case, a pressure maximum is formed inside the disk (at $R_{2}$) and plasmas are "pushed" towards the BH, forming
a `pressure-driven' flow and resulting in a `disk-like' structure on the equatorial plane near the horizon. On the contrary, if $\ell(R)<\ell_{\rm ms}$,
the sub--Keplerian plasmas fall into the BH because of  the nature of their sub-Keplerian augular momentum (`viscous-driven').
%inside $R_{\rm ISCO}$ 
The plasmas are able to fall closer to the axis (i.e., higher latitudes of the horizon), forming  a `quasi-spherical' geometry near the horizon.  These two kinds of accretion flow geometry near the horizon, \textit{disk--like} and \textit{quasi--spherical}, was first recognized by \citet{abr81}.

Figure 1 schematically depicts the geometric and dynamic properties of different accretion disks
and the extraction of the BH rotational energy  when
different accretion flow geometries near the horizon is realized. 
If the magnetosphere near the horizon is filled with plasmas, 
those plasma can load onto the large--scale field lines and reduce the electromagnetic 
extraction of the BH energy \citep{tak90}.
In this letter, by assuming that a relativistic jet are launched when the BH energy is extracted outward, we investigate the formation of BH relativistic jets when the BH is surrounded by  different type of disks listed in Table 1.
This work is an extension of \citet{pu12}, hereafter PHC, in which the formation of relativistic jets at
low accretion rate (e.g. when $\dot{m}\lesssim 0.1$ ) is considered. 
It is found that, ergospheric jets can preferentially take place not only when
the BH is surrounded by 
a \textit{combined disk}, as suggested in PHC,
but also when the BH is surrounded by a \textit{slim disk}. 
We refer the two types of jets as `EJC' and `EJS', respectively (see section 3).
We conclude that the characteristic accretion rate of a EJC and  a EJS
are $\dot{m}<0.01$ and  $\dot{m}>0.01$,
respectively, and
that  the
EJS is in general more powerful and faster than the EJC for the same central BH mass and BH spin. 
The difference between these two types of jets may result in 
the observed dichotomy of of FR I 
and FR II galaxies.

\section{Ideal MHD Flow}
In an axis--symmetry and stationary BH magnetosphere, the total energy of the ideal MHD flow, $E$, is conserved along a specific field line (see, e.g., PHC for details). We can separate $E$ into the plasma part (the first term) and electromanetic part (the second term),
\begin{equation}
E=-\mu u_{t}+\frac{\Omega_{\mathrm{F}}}{4\pi\eta}B_{\phi}\;,
\end{equation}
where $\mu$ is the relativistic specific enthalpy, $u_{t}$ is convariant time component of the four---velocity,
$\Omega_{\mathrm{F}}$ is the angular velocity of the field, $\eta \propto n u^{r}/ B_{r}$
 is the particle
flux per unit flux tube, $n$ is the proper number density,  $u^{r}$ is the radial
component of the four--velocity
and $B_{\phi}$ and $B_{r}$ is  the toroidal and radial field observed by
distant observer. 
Similarly,
the total outward energy flux $\mathcal{E}^{r}$  has  the plasma part, $\mathcal{E}_{\mathrm{em}}^{r}$,
and the electronmagnetic 
part, $\mathcal{E}_{\mathrm{em}}^{r}$,
\begin{equation}
\label{eqn1}
\mathcal{E}^{r}=nEu^{r}=\mathcal{E}_{\mathrm{em}}^{r}+\mathcal{E}_{\mathrm{plasma}}^{r}\;.
\end{equation}

A plasma is accelerated along the large-scale magnetic field lines inwards (or outwards)
if the gravitational force is larger (or less) than the magnetocentrifugal forces. 
The conservation of $\mathcal{E}^{r}$ near the separation region,
where the magnetocentrifugal forces 
balance with the gravitational force, requires the solution of a 
powerful outflow (i.e. relativistic jet) to that of
an inflow  with a positive energy flux ($\mathcal{E}^{r}>0$).
As discussed in $\S$3.6 of PHC, for the inflow,
$\mathcal{E}_{\mathrm{em}}^{r}>0$ and
$\mathcal{E}_{\mathrm{plasma}}^{r}<0$  are  generally expected; that is, 
the magnetic field lines contribute to the extraction of the
rotational energy of a BH, while the plasma plays an opposite role and reduces the extraction.
Therefore,
an MHD flow can extract the BH energy (accordingly a {\em ergospheric jet} is launched), only when 
the magnetosphere is magnetically dominated,
$|\mathcal{E}_{\mathrm{em}}^{r}|>|\mathcal{E}_{\mathrm{plasma}}^{r}|$.
The extraction of the rotational energy by the MHD flow is described by the `MHD Penrose process' \citep{tak90}.
When $\mathcal{E}_{\mathrm{plasma}}^{r}$ is
negligibly small compared to $\mathcal{E}_{\mathrm{em}}^{r}$ (i.e., a nearly vacuum environment),
the force-free limit becomes a good approximation.
The electromagnetic extraction of the rotational energy of a BH (by large-scale, hole-threading field lines) is 
described by the `Blandford-Znajek process' \citep{bla77}.
\\
\\
\\
\\
\begin{figure}[ht]
\label{fig1}
%\epsscale{.5}

(a)

\includegraphics[width=4cm]{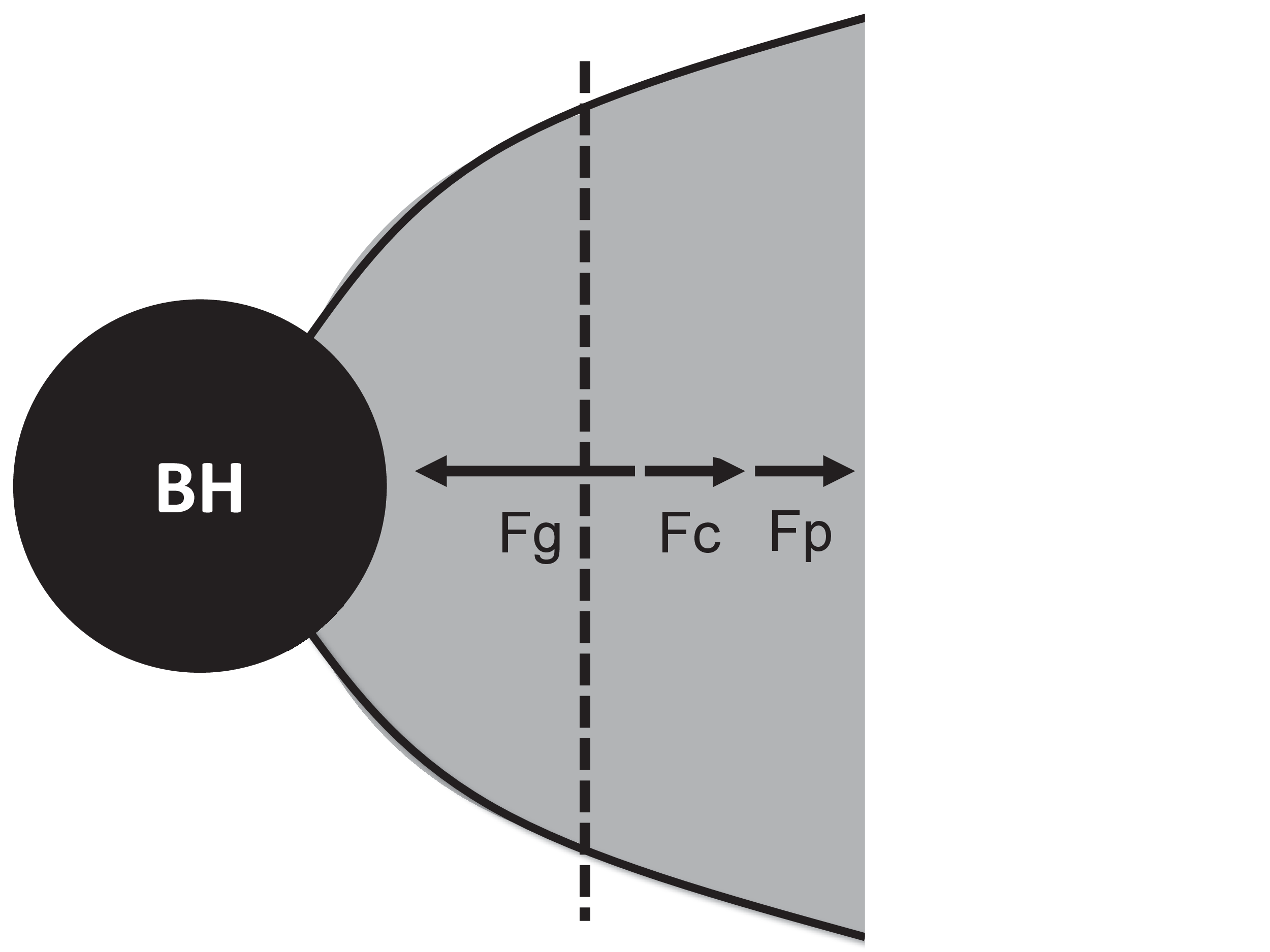}\includegraphics[width=4cm]{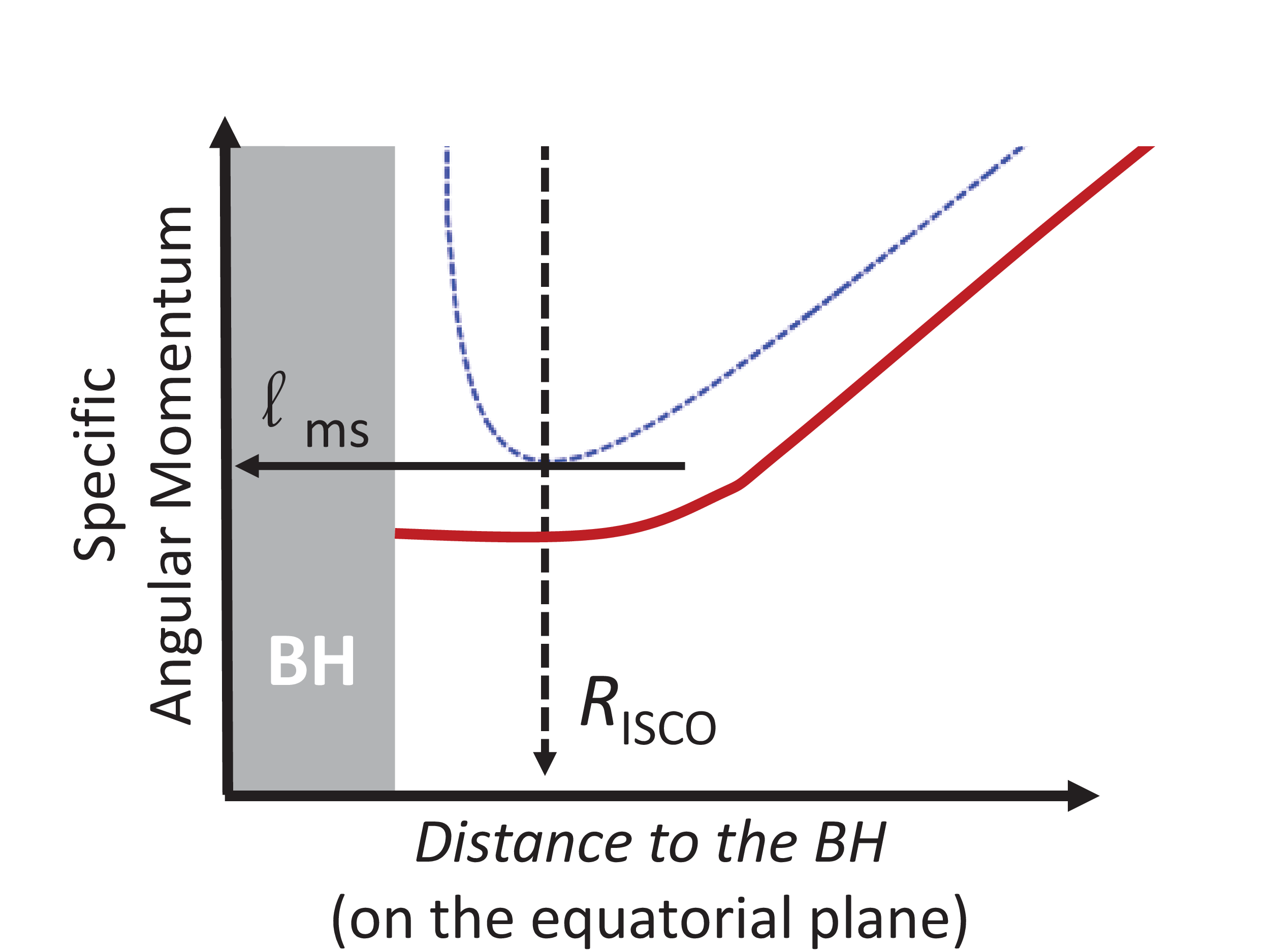}

(b)

\includegraphics[width=4cm]{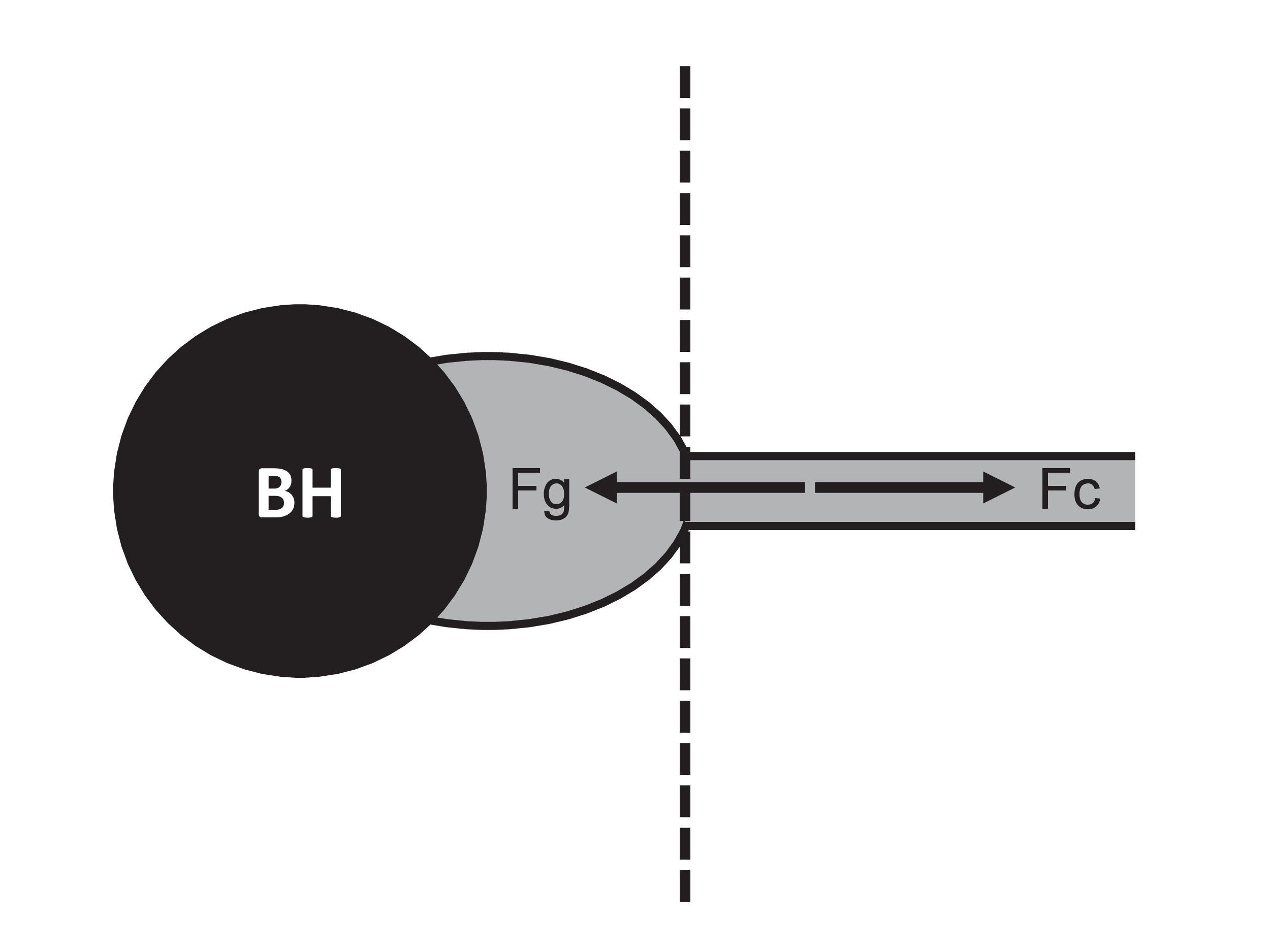}\includegraphics[width=4cm]{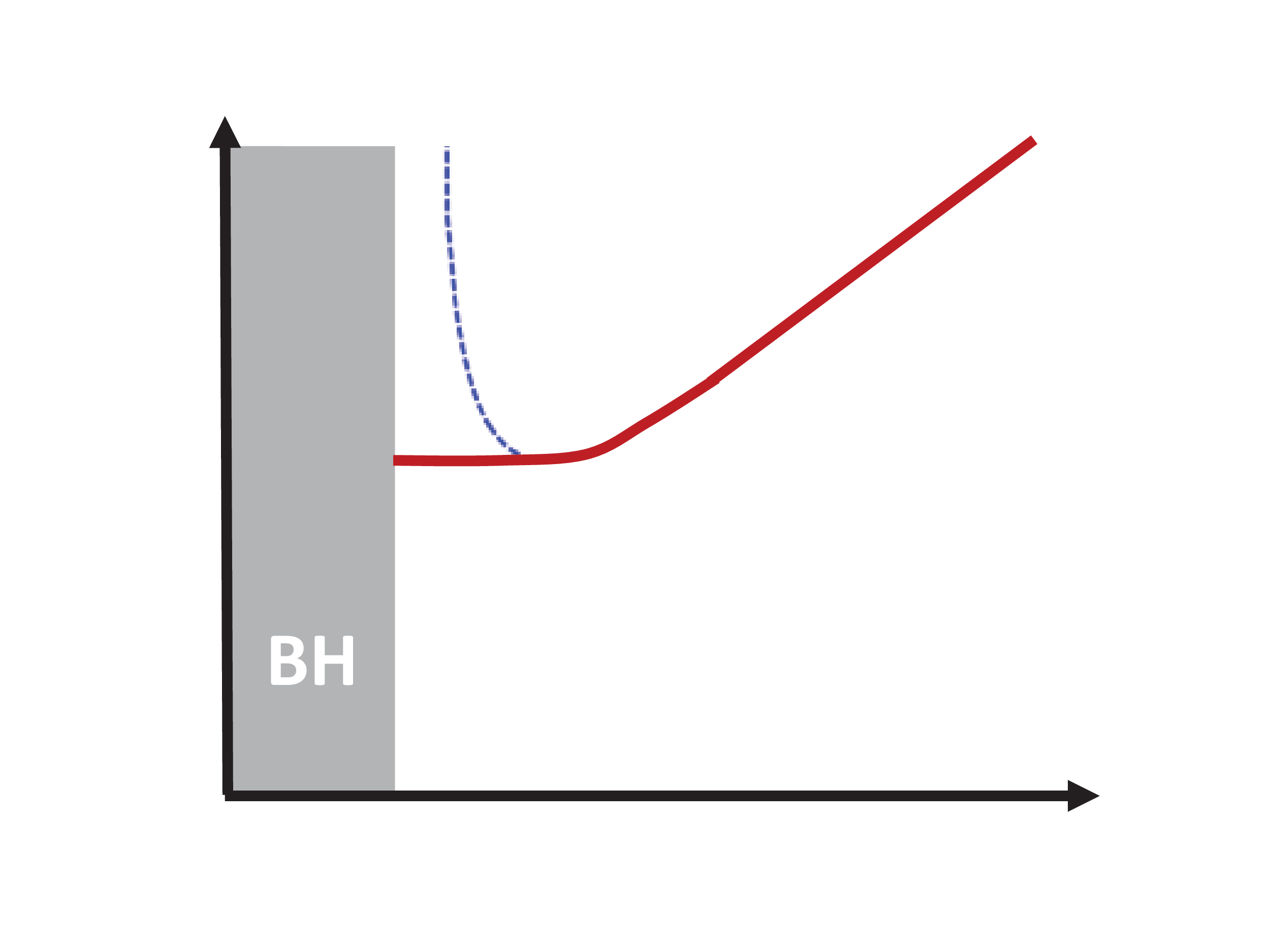}

(c)

\includegraphics[width=4cm]{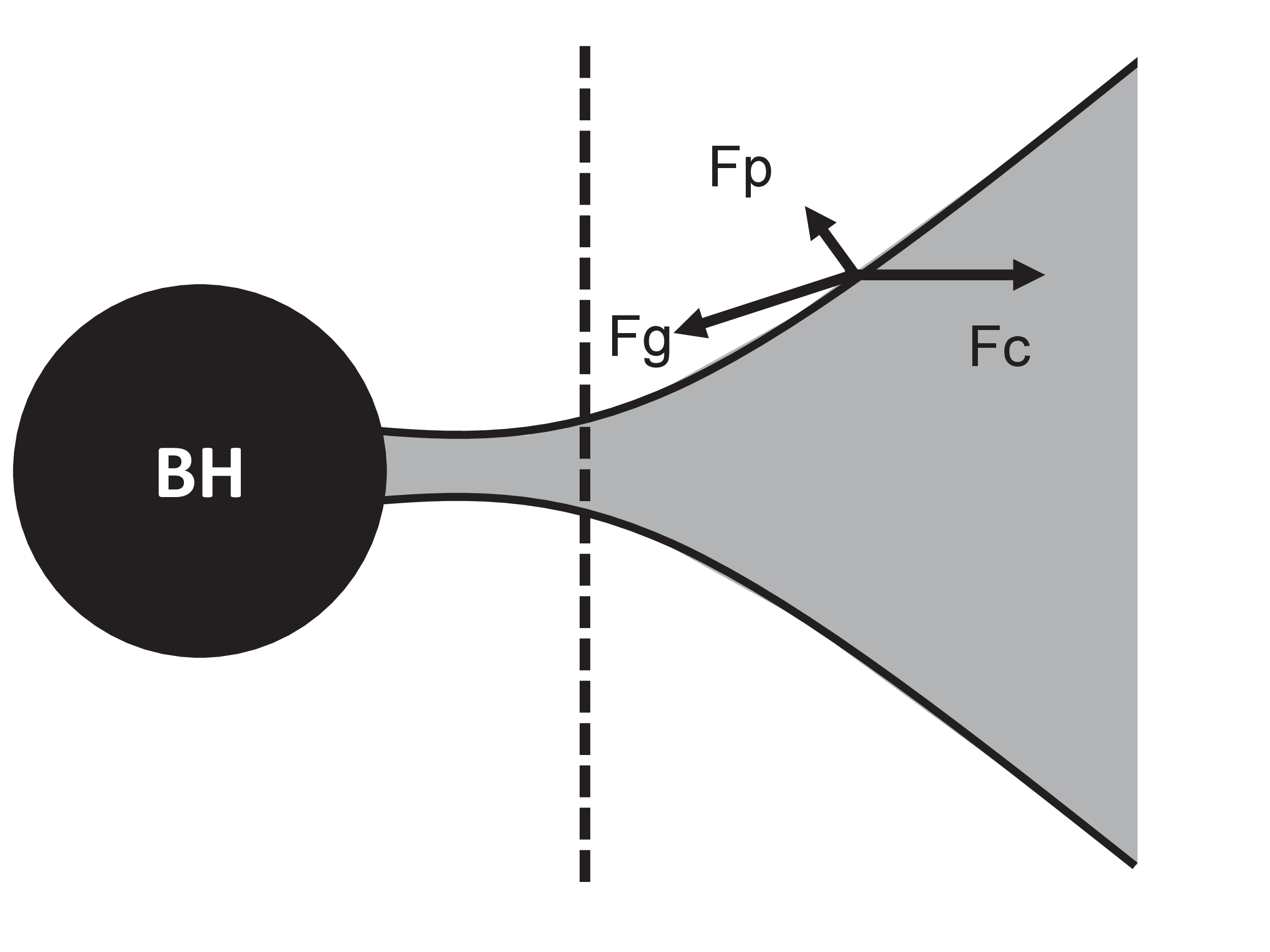}\includegraphics[width=4cm]{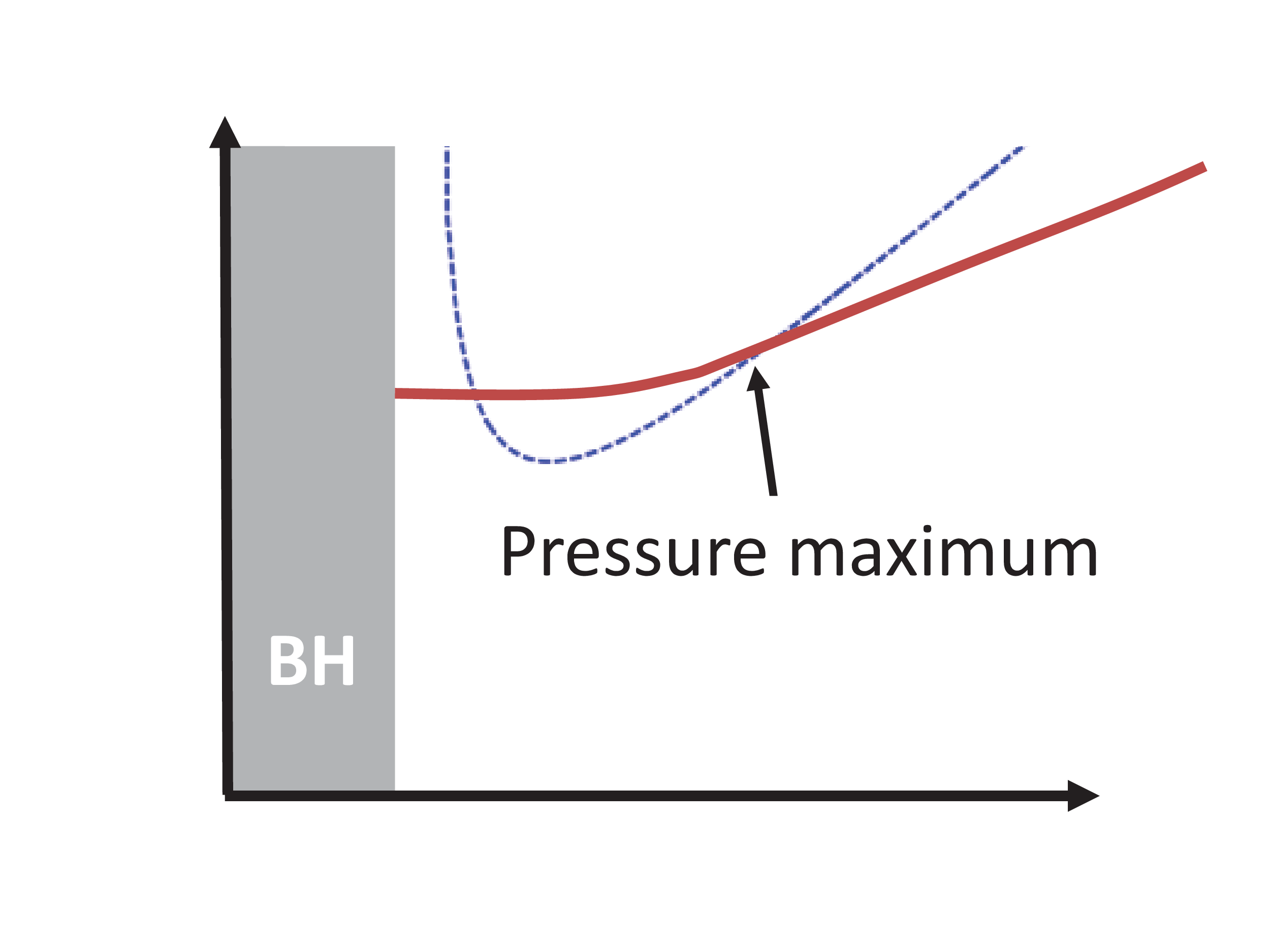}

(d)
\begin{center}
\includegraphics[width=6cm]{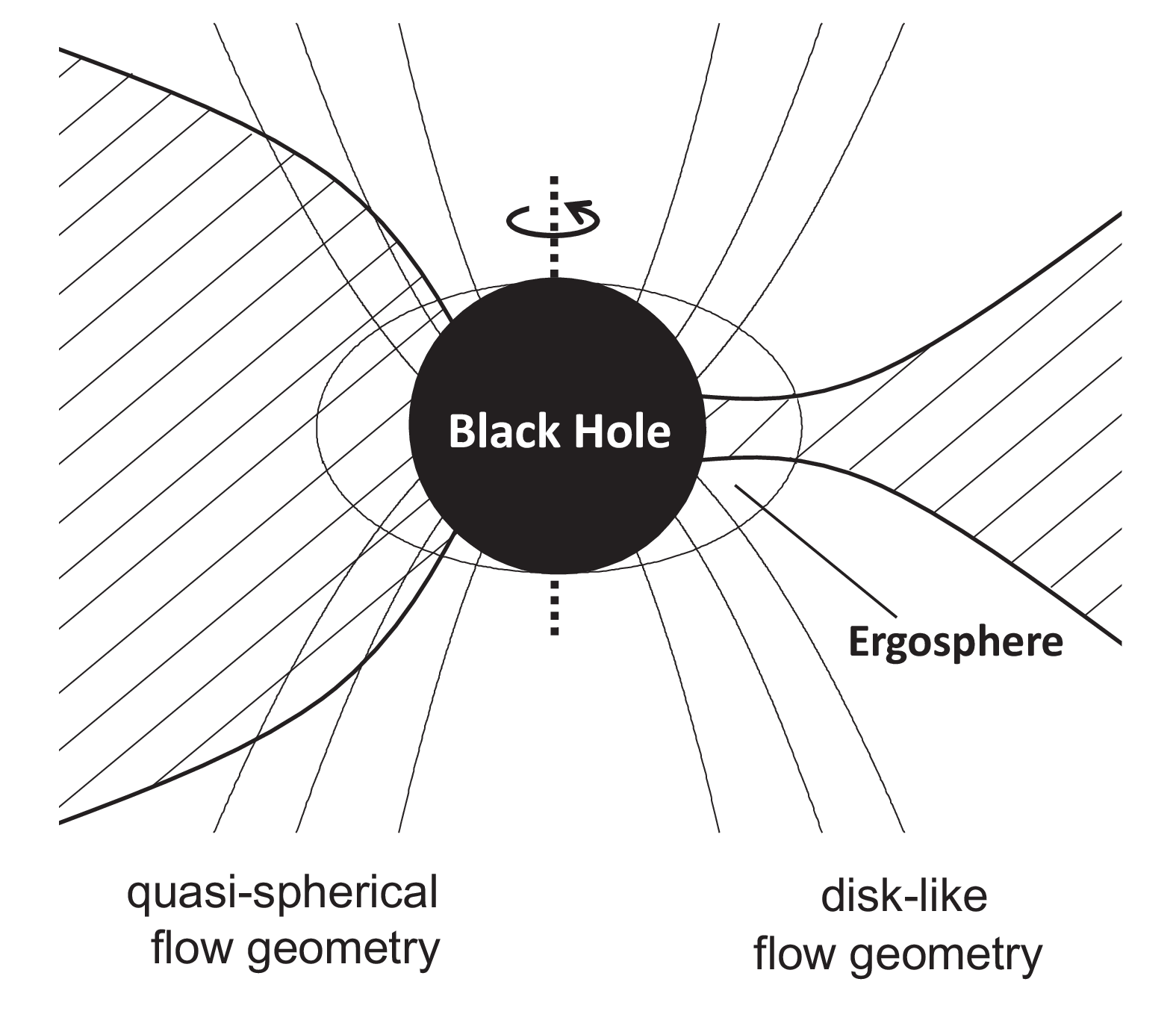}
\end{center}
%\plotone{fig1.ps}
\caption{(a)Illustration plot (left) and representative angular momentum profile (right) of a ADAF. Dominating force are shown: $F_{\rm g}$, force due to the gravity of the BH;  $F_{\rm c}$, force due to the rotation of the disk; $F_{\rm p}$, force due to the pressure gradient inside the disk. The dashed line represents the location of $R_{\rm ISCO}$. The specific  angular momentum of the disk is shown in red solid curve while the Keplerian value is shown in dashed bule curve. (b)Same as (a), but of a thin disk. (c) Same as (a), but of a slim disk. 
(d)Extraction of BH rotational energy by a large--scale magnetic field under different accretion flow geometries near the BH horizon (cf. Figure 2 of \citet{abr81}). The field lines are indicated by solid lines and the accreting plasma
are indicated by the shaded region (not to scale). 
%The static limit is denoted by the solid lines that enclose the ergosphere.  
(left)Quasi-spherical flow geometry. Such geometry is resulting from a viscous-driven acrretion process, such as an ADAF or a thin disk. 
The electromagnetic extraction of the BH rotational energy by the field lines is reduced, because the inflowing plasmas along the large--scale field lines bring 
the  rest--mass energies onto the BH.
For the thin disk case, the
depicted region here is corresponding to a much inner region than $R_{\rm ISCO}$. 
(right)Disk-like flow geometry. Such geometry is resulting from a pressure-driven accretion process, such as a slim disk.} 
%(Color version of this figure is available in the online journal)} 
\end{figure}

\section{Two Types of Relativistic Jet} 
In general,  
the strength of a large-scale magnetic field strength\footnote{We should stress here that the estimation of \textit{large-scale field} should be different from the estimation of the \textit{local field}, which can be computed by the plasma beta and the gas pressure inside the disk.
Numerical simulations, e.x. \citet{mck12} and the references therein, show that the accumulated field strength near the BH 
%is not necessary has an upper limit of the local field strength; instead, it 
can even reach to an equipartition level and 
supports the accretion disk.}
can be parameterized by the ratio, $\varepsilon^{2}$, 
of the gravitational binding energy of the disk at radius  $R$
to the large-scale magnetic field energy inside  $R$ (see also PHC). That is,
\begin{equation}
B(\varepsilon,R,\Sigma(R))=\varepsilon\sqrt{2\pi GM_{\mathrm{BH}}\Sigma/R^{2}}\;,
\end{equation}
where $G$ is the gravitational constant and $M_{\mathrm{BH}}$ the mass of the central
BH. 
When $\varepsilon=1$, $B$ reaches a equipartition value at the radius $R$.

By examining whether the BH rotational energy can be extracted by the MHD flow 
via the MHD Penrose process  
 when the BH is surrounded by
an ADAF, a combined disk or a thin disk,
PHC concluded that
the jet is most likely launched when the surrounded disk is of 
a combined disk type among all the cases.
Because the surface density of the 
outer thin disk is much larger than that of the inner ADAF for a combined disk,
the inner ADAF can therefore `receive' 
more large-scale fields from the outer thin disk than what a single ADAF would capture (see equation [3]). Therefore, the quasi-spherical plasma provided by the inner ADAF of a combined disk can be loaded on 
relatively stronger field lines.
Relativistic jets can be finally initiated when
the magnetosphere  becomes  magnetically
dominated. This situation can be realized when
the transition radius, $R_{\rm{tr}}$, where the outer thin disk transits to the inner ADAF,  closes enough 
to the BH. 
For the convenience of discussion, hereafter we refer this type  of jets
as \textit{ergospheric jet from the center of a  combined disk (EJC)}.

Ergospheric jet can also be preferentially launched when the BH
is surrounded by a slim disk.
In this case,  plasmas enter the BH through a disk-like structure on the equatorial plane and  most of the region near the horizon remains nearly vacuum (Figure 1b).
Therefore, the Blandford--Znajek process can efficiently operate.  
Hereafter we refer this type of jets as
\textit{ergospheric jet from the center of a slim disk (EJS)}.
The power and the speed of the JEC and the  EJS are discussed in what follows.

\subsection{Jet Power} 

\begin{figure}[ht]
\label{fig2}
%\epsscale{.6}
\plotone{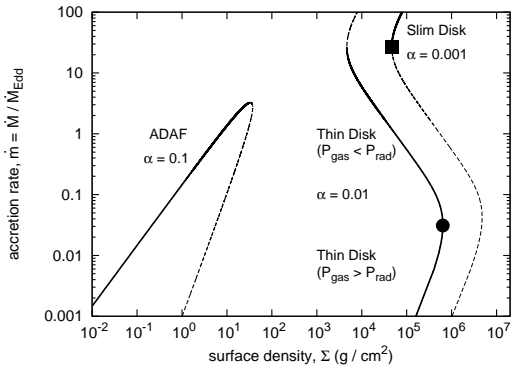}
\caption{Thermal equilibira for accretion disks at $R/R_{g}=10$ when $M_{\rm{BH}}=10^{9} M_{\odot}$. 
The solid curves, from left to right,
shows the ADAF soltuion with $\alpha=0.1$, thin disk solution with $\alpha=0.01$ and slim disk solution
with $\alpha=0.001$. The thin disk solution become unstable when the radiation pressure, $P_{\rm rad}$, dominates
the gas pressure, $P_{\rm gas}$.
The filled circle indicates the maximum surface
density, $\Sigma$, of a thin disk with a fixed $\alpha$.
The maximum $\Sigma$ is used to evaluate the
the maximum large-scale field
strength that a combined disk can reach (see text for detail). Note that 
the pseudo-Newtonian potential is adopted here. In a full relativistic computation, the 
upper turning point of the "S" curve, which is indicated by the filled square and corresponds
to   
the minimum $\dot{m}$ 
of a slim disk,
has a value $\dot{m}\gtrsim0.3$ \citep{sad09,abr10}, 
unlike a large value ($\dot{m}\gg1$) here. }
\end{figure}

\begin{figure}[ht]
\label{fig3}
%\epsscale{.6}
\plotone{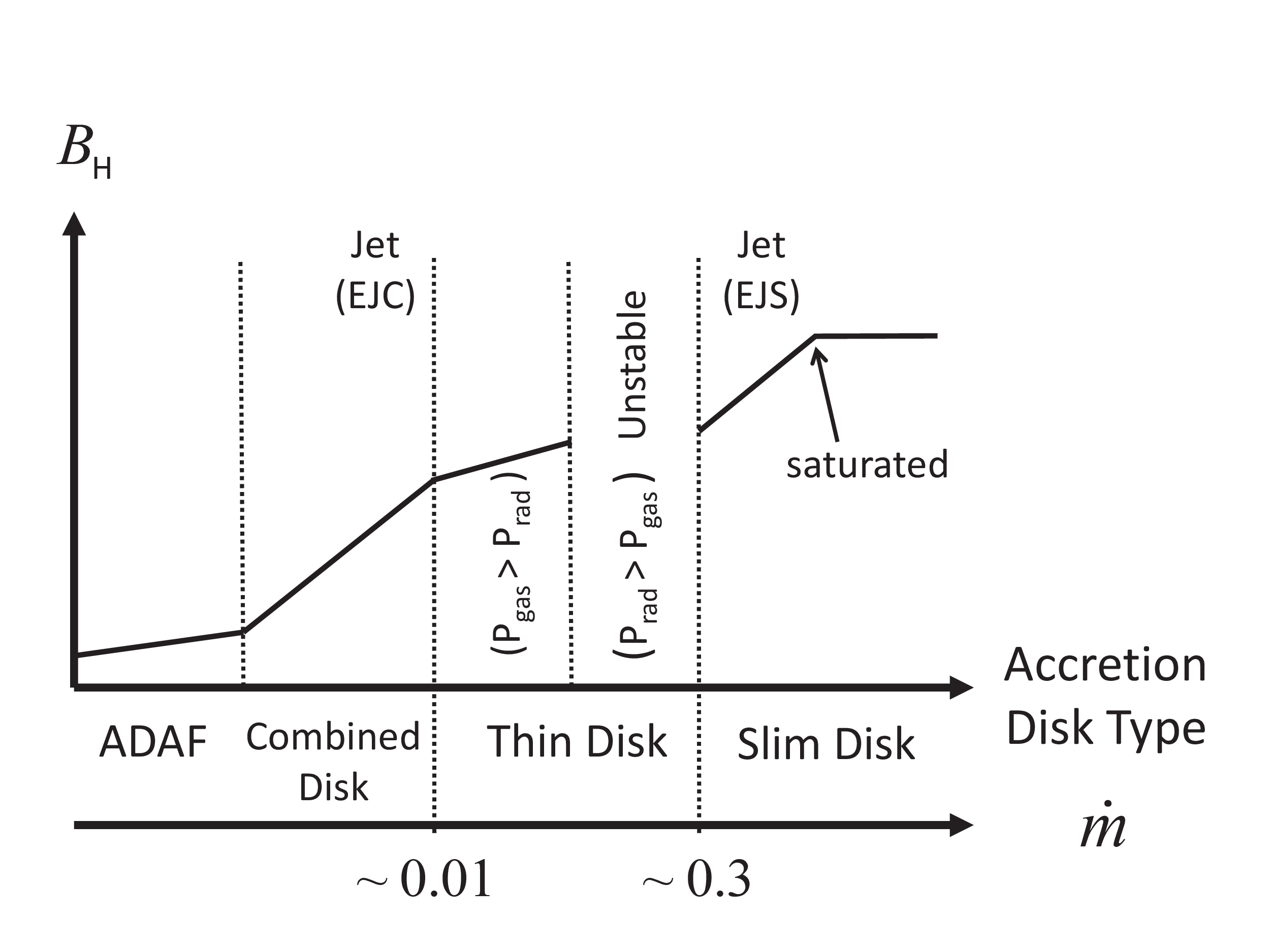}
\caption{Relative large-scale magnetic field strength near the BH, $B_{\rm{H}}$, as a function of the accretion--disk type (or, equivalently, $\dot{m}$) for a roughly fixed $\varepsilon$, according to Equation (3). 
The increase of the field strength with increasing  
$\dot{m}$ in a slim disk
levels off at a
saturated value. 
The resulting magnetically dominated magnetosphere when
the BH is surrounded by a combined disk or a slim disk, can respectively result in 
two types of jets, EJC and EJS. See text for more details.}
\end{figure}

\begin{figure}[ht]
\label{fig4}
%\epsscale{.6}
\plotone{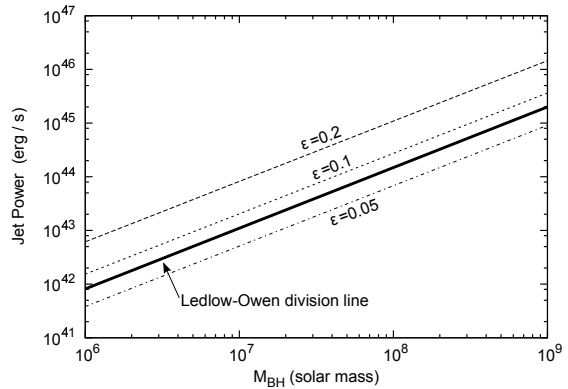}
\caption{Maximum power of the EJC as a function of BH masses with  discrete values of $\varepsilon$'s. 
The long--dashed, short--dashed, and dash--dotted lines represents the case of $\varepsilon=$0.2, 0.1 and 0.05, respectively.
The power
is computed by Equation (5), after the maximum value of  $B_{\rm{H}}$ is estimated by
Equation (3). The spin of the BH, $j$, is assumed to be $0.6$. Since the mass loading
or a smaller spin can further lower the power,  the power of EJC will appear in
the lower part of these maximum power lines.
The   FR I/ FR II division line found by \citet{led96} is represented by 
the solid thick line. Most of FR I (FR II) sources are located above (below) this division line.}
\end{figure}

The total extracted power from a rotating BH can be computed by $P\equiv\int_{R_{H}}\mathcal{E}^{r}dS$, where $R_{H}$ is the radius of the black hole horizon. Similar to Equation (1), we can further divide the power into the electromagnetic part, 
$P_{\rm{em}}$, and the plasma part, $P_{\rm{em}}$, that is, $P=P_{\rm{em}}+P_{\rm{plasma}}$, where
\begin{equation}
P_{\rm{em}}\equiv\int_{R_{H}}\mathcal{E}_{\rm{em}}^{r}dS\,\,,
\end{equation}
and
\begin{equation}
P_{\rm{plasma}}\equiv\int_{R_{H}}\mathcal{E}_{\rm{plasma}}^{r}dS\,\,.
\end{equation}
The mathematical similarity of the 
idea MHD assumption and the  force-free assumption
guarantees that $P_{\rm{em}}$ is identical to the output power of 
Blandford--Znajek  process:
\begin{equation}
P_{\rm{em}}=\frac{1}{32}\omega_{F}^{2}B_{\rm{H}}^{2}R_{\rm{H}}^{2}j^{2}c\,\,,
\end{equation}
where $j\equiv J/J_{max}=J/(GM^{2}/c)$, $J$ the angular momentum
of the BH,
 $B_{\rm{H}}$ the large-scale hole-threading field strength,
 and $\omega_{F}^{2}\equiv\Omega_{F}(\Omega_{H}-\Omega_{F})/\Omega_{H}^{2}$, $\Omega_{\rm{H}}$ the rotational 
velocity of the BH.
Assuming  $\omega_{F}=1/2$, which maximize $P_{\mathrm{em}}$,
we have, 
\begin{equation}
P_{\mathrm{em}}=\frac{1}{32}\frac{1}{4}B_{\rm{H}}^{2}\left[1+\sqrt{(1-j^{2})}\right]^{2}(\frac{GM_{\rm{BH}}}{c^{2}})^{2}\, j^{2}c\,\,.
\end{equation}
In slim disk case, the accreting plasma, which contributes to $P_{\rm{plasma}}$, stay on the equatorial plane. As a result, there is only tiny amount of plasma loading onto the large--scale magnetic field lines that thread the horizon at different latitudes. The jet power
of a EJS can be therefore estimated by Equation (7), namely that
$P_{\rm{EJS}}\approx P_{\rm{em}}$.  
On the other hand, 
because the plasma with quasi-spherical geometry near the horizon can  be loaded onto the field lines and can
further reduced the power by the amount of $P_{\rm{plasma}}<0$,  Equation (7) represents 
the \textit{maximum} power of a EJC.

By Equation (3), $B_{\rm{H}}$ can be computed if  $\varepsilon$ and
$\Sigma$ are known. 
Although the determination of $\varepsilon$ is unclear, it should
in general has a value $1>\varepsilon>0$.
Therefore, the changing of $B_{\rm{H}}$ at different accretion rate should  
mainly be determined by 
the variation of the 
surface density of the disk, $\Sigma (\dot{m})$, because it will be higher variable than 
$\varepsilon (\dot{m})$.
Here we assume that the large-scale, disk-threading fields can be dragged inward
by accretion flow for \textit{all} types of disks \citep{nar03,spr05,rot08} and hence 
$\varepsilon>0$ is always satisfied.
Note that, in a slim disk,  it is 
the pressure--driven nature (instead of the viscosity) that determines the advection motion of the plasma near the BH. 
Therefore, although the the magnetic Prantdl number is expected to be small due to a relative
tiny value of $\alpha$ compare to other types of disk (Table 1),  
the field strength near the horizon can be further enhanced when the disk transits form a thin disk to 
a slim disk, by further pushing a large-scale, disk-threading field lines inward.

Figure 2 presents typical BH accretion disc solution profiles   on the $\Sigma$ -- $\dot{m}$ plane,
following the model in \citet{abr95} and \citet{che95}.
The solid curves represent the solutions of a ADAF, a thin disk and a slim disk with their typical
viscosity (see figure caption).
With the relation of $\Sigma$ -- $\dot{m}$ and Equation (3),
relative values of  $B_{\rm{H}}$ 
as a function of $\dot{m}$
are qualitatively presented in Figure 3.
The field strength gradually increases with increasing $\dot{m}$ when the disk is either an
ADAF or a thin disk because of 
the increase of $\Sigma$, where the radiation-pressure dominated thin disk solution is not of interest here because the disk become unstable in that case.
For a combined disk,  $B_{\rm{H}}$
is essentially described by the surface density of the disk at the transition radius, $R_{\rm{tr}}$, because the 
outer thin disk has a much larger density than that of the inner ADAF.
The  decrease of $R_{\rm{tr}}$
with increasing $\dot{m}$ results in a 
rapid growth of $B_{\rm{H}}$ with $\dot{m}$ 
(see also the Figures 1a and 1e of PHC).
For a slim disk,  although 
$B_{\rm{H}}$ 
can be further enhanced due to the pushed-in large-scale, disk-threading field lines,
 the increase of $B_{\rm{H}}$ is expected to 
saturate eventually. This is because the fields 
diffuse outward outside the pressure-driven region (that is, outside the 
super-Keplerian part of the disk).

The transition radius of a combined disk continuously decreases with increasing $\dot{m}$ before the entire
disk become a thin disk.
Previous studies \citep{hon96,man00} show that $R_{\rm{tr}}$ can reach down to
 $\sim10R_{g}$.
For illustration purpose, assuming a modest BH spin, $j=0.6$,
we compute the maximum value of $B_{\rm{H}}$ for a combined disk 
 by using the maximum $\Sigma$ of the (outer) thin disk solution at $R=10R_{g}$ (the filled circle in Figure 2) and 
calculate the resulting \textit{maximum} power of a EJC by equation (3).
The jet power for three different $\varepsilon$'s
are plotted on the $M_{\rm{BH}}$ -- $P$ plane (Figure 4).
Noting that that the plasma contribution, $P_{\rm{plasma}}<0$, as well as a smaller BH spin, will reduce the power
of EJC,
we find these jet powers represent the upper limits.
In other words, the power of a EJC will appear below these lines.
On the contrary, for the same parameters (BH mass, BH spin, $\varepsilon$),
sources with EJSs will appear above the lines
because of a larger $B_{\rm{H}}$ (Figure 3) and a negligible
 $P_{\rm{plasma}}$ in the force-free limit.

\subsection{Jet Speed}
By denoting the quantities of the inflow ($u^{r}<0$) and outflow ($u^{r}>0$)  
with the subscript "in" and "out", respectively,
the ratio of the jet speeds of a EJC and a EJS can be estimated as follows:
At large distances, the jet Lorentz factor, $\Gamma$, can be defined by
\begin{equation}
 \Gamma=\frac{E_{\rm{out}}}{\mu}\;. 
\end{equation}
In addition, near the separation region of the inflow and the outflow, the conservation of $\mathcal{E}^{r}$ 
gives, 
\begin{equation}
E_{\rm{out}}
=\frac{n_{\rm{in}}u^{r}_{\rm{in}}}{n_{\rm{out}}u^{r}_{\rm{out}}}E_{\rm{in}}\;. 
\end{equation}
%Therefore,
%\begin{equation}
% \frac{\Gamma^{\rm{EJC}}}{\Gamma^{\rm{EJS}}}
% =\frac{E_{\rm{out}}^{\rm{EJC}}}{E_{\rm{out}}^{\rm{EJS}}}
%=\frac{(n_{\rm{out}}u^{r}_{\rm{out}})^{\rm{EJS}}}
%{(n_{\rm{out}}u^{r}_{\rm{out}})^{\rm{EJC}}}\;. 
%\end{equation}
Thus, for fixed BH  mass and  spin, the ratio of the power of the EJC and the EJS, $\chi$, can be written as,
\begin{equation}
 \chi\equiv\frac{\Gamma^{\rm{EJC}}}{\Gamma^{\rm{EJS}}}
% =\frac{E_{\rm{out}}^{\rm{EJC}}}{E_{\rm{out}}^{\rm{EJS}}}
\approx\frac{(n_{\rm{out}}u^{r}_{\rm{out}})^{\rm{EJS}}}
{(n_{\rm{out}}u^{r}_{\rm{out}})^{\rm{EJC}}}
\times\frac{(B_{r} B_{\phi})^{\rm{EJC}}}
{(B_{r} B_{\phi})^{\rm{EJS}}}\;, 
\end{equation}
by the help of Equation (1),
where the superscript "EJC" ("EJS") denotes the parameter of EJC (EJS).
Because
both   $(n_{\rm{out}}u^{r}_{\rm{out}})^{\rm{EJS}}/(n_{\rm{out}}u^{r}_{\rm{out}})^{\rm{EJC}}<1$ 
and  $(B_{r} B_{\phi})^{\rm{EJC}}/(B_{r} B_{\phi})^{\rm{EJS}}<1$ hold, we obtain $\chi\ll 1$.

\section{Discussion}
The characters of EJC and EJS offer the key to an understanding of the FR I/ FR II dichotomy.
The   FR I/ FR II division line found by \citet{led96}, which is shown by the thick solid line in Figure 4,
is roughly consistent with the maximum power that a EJC can reach. 
It is, therefore, reasonable to suppose that the division reflects the two
types of relativistic jets. That is,
most FR I (or FR II) galaxies are associated with a EJC
(or a EJS). 
Noting that a EJC has a slower speed than a EJS,  why 
 FR I and FR II galaxies have a edge-darken and a edge-brighten morphology, respetively, can be consistently understood as a result of different jet velocities \citep{bic85}.

\citet{ghi01} found that  the
accretion rate of FR I and FR II galaxies can be separated 
at $\dot{m}\approx0.01$.
Similar division of the accretion rates is also indicated
for  BL Lacs and radio quasars 
(which are believed to be  `face-on' FR I and FR II galaxies, respectively) \citep{xu09,ghi11},
and for radio galaxies of `high-excitation' and `low-excitation' \citep{bes12}.
The corresponding $\dot{m}$ of EJC ($\dot{m}<0.01$) and EJS ($\dot{m}>0.01$) are consistent with the above findings.

The major drawback in our calculation is that,
the Pseudo-Newtonian potential \citep{pac80} is adopted when we use
used the disk  model of \citet{abr95} and \citet{che95}.
We should notice
a full-relativistic model (e.g. \citet{sad09}), could significantly change the range
of $\dot{m}$ of 
the slim disk solution  (see captions of Figure 2). Nevertheless, the surface
density of the disk solution will not be changed dramatically  \citep{sad09,abr10,sad11}.
We therefore expect that the results will be qualitatively unchanged when general--relativistic corrections are taken into account.
The effect of the BH spin
on the jet power,
which is beyond the present model, can be investigated in the future by
incorporating general relativistic corrections.

%\clearpage{}

\acknowledgments
\label{ack}
H.-Y. thanks Daniel Chun-Cheng Lin for suggestions of figure plotting.
This work was supported by the National
Science Council (NSC) of Taiwan under the grant NSC
99-2112-M-007-017-MY3,
NSC 100-2112-M-007-022-MY3,
and the Formosa Program between NSC and Consejo
Superior de Investigaciones Cientificas in Spain administered under
the grant NSC100-2923-M-007-001-MY3.

%\clearpage{}

\label{reference}

\end{document}